\documentclass[conference]{IEEEtran}
\IEEEoverridecommandlockouts
\usepackage[letterpaper, top=0.8in, bottom=1.04in, left=1in, right=1in]{geometry}
\usepackage{pgfplots}
\pgfplotsset{compat=1.18}
\usepackage{graphicx}
\usepackage{array}
\usepackage{cite}
\usepackage{amsmath,amssymb,amsfonts}
\usepackage{algorithmic}
\usepackage{textcomp}
\usepackage{xcolor}
\usepackage{hyperref}
\usepackage{eso-pic}

\begin{document}

\title{A Deployment-Aware Feasibility Framework for Machine Learning-Based IoT Intrusion Detection Across Edge, Fog, and Cloud Architectures\\
}
\author{
\IEEEauthorblockN{Shaker Nawasra}
\IEEEauthorblockA{
School of Information Security \& Applied Computing \\
Eastern Michigan University \\
Ypsilanti, USA \\
snawasra@emich.edu
}
\and
\IEEEauthorblockN{Munther Abualkibash}
\IEEEauthorblockA{
School of Information Security \& Applied Computing \\
Eastern Michigan University \\
Ypsilanti, USA \\
mabualki@emich.edu
}
}
\maketitle

\begin{center}
\small
DOI: \href{https://doi.org/10.1109/CSNT69054.2026.11502111}
{10.1109/CSNT69054.2026.11502111}
\end{center}

\AddToShipoutPictureFG*{
\AtPageLowerLeft{
\hspace{0.7in}
\raisebox{0.35in}{
\parbox{3.3in}{
\footnotesize
© 2026 IEEE. Personal use of this material is permitted.
Permission from IEEE must be obtained for all other uses, including
reprinting/republishing this material for advertising or promotional
purposes, creating new collective works, for resale or redistribution
to servers or lists, or reuse of any copyrighted component of this
work in other works.
}}
}
}

\begin{abstract}
The rapid growth and heterogeneity of Internet of Things (IoT) environments have exposed fundamental limitations in traditional rule-based and signature-based intrusion detection systems. This paper presents a quantitative deployment-aware analysis of machine learning (ML)–based intrusion detection approaches across edge, fog/gateway, and cloud architectures. Unlike prior surveys that primarily emphasize detection accuracy, this work defines representative quantitative deployment capability envelopes extracted from experimental and system-level studies and introduces a structured Deployment Feasibility Score (DFS) model. The proposed framework maps ML techniques to architectural layers based on computational demand, memory footprint, and latency sensitivity using a weighted ordinal scoring mechanism. The analysis demonstrates that lightweight statistical and linear models are most suitable for edge deployment, ensemble and clustering-based methods align with fog/gateway environments, while deep and optimization-driven models are best suited for cloud infrastructures. By formalizing deployment feasibility through quantitative grounding and structured evaluation, this work provides practical guidance for selecting intrusion detection solutions under real-world architectural constraints, supporting more informed and deployment-conscious IoT security design.
\end{abstract}

\begin{IEEEkeywords}
Internet of Things (IoT), intrusion detection systems, machine learning, deployment feasibility, edge computing, fog computing, cloud computing.
\end{IEEEkeywords}

\section{Introduction}
IoT devices typically operate under strict resource constraints, including limited computational power, memory, and energy availability, and often rely on lightweight operating systems and communication protocols. IoT devices primarily collect data from the physical environment through embedded sensors and are frequently deployed at scale with minimal manual configuration. Once powered on, many IoT devices automatically connect to the network, use default or preconfigured settings, and start sending data without manual tuning. As a result of their constrained hardware capabilities, simplified software stacks, and the frequent prioritization of functionality over security during rapid deployment, IoT devices are inherently vulnerable to a wide range of cyber threats \cite{en16083465}. 

\begin{figure}[h]
\centering
\includegraphics[width=0.85\linewidth]{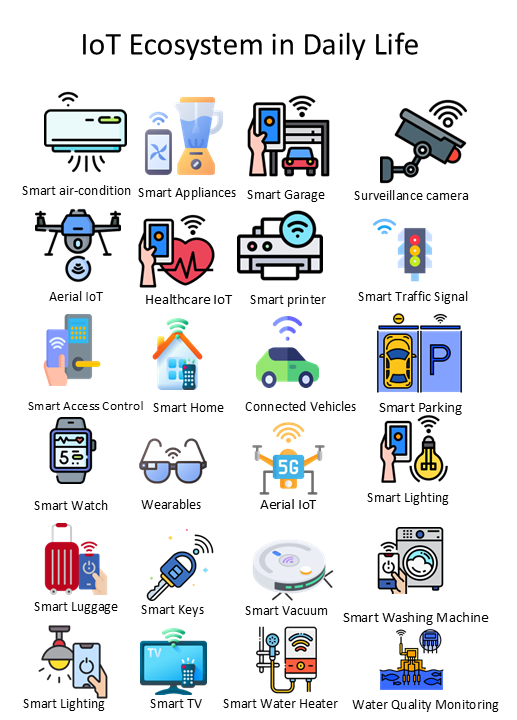}
\caption{IoT Ecosystem in Daily Life.}
\label{fig:iot_ecosystem}
\end{figure}

IoT devices vary widely in their operating systems, computing power, and functions, as illustrated in Fig.~\ref{fig:iot_ecosystem}, making it difficult to design a single security system that can effectively protect all different kinds of IoT devices.

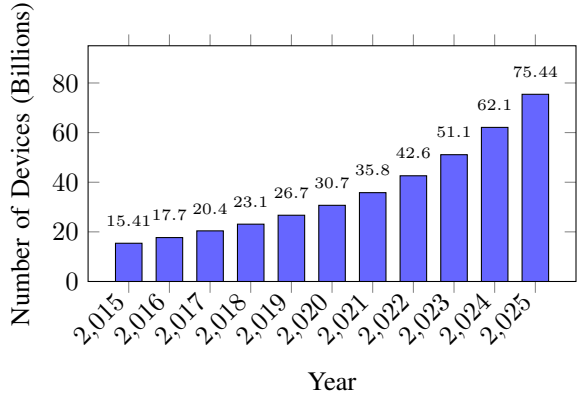
\begin{figure}[h]
\centering
\begin{tikzpicture}
\begin{axis}[
   width=\columnwidth,
   height=4.7cm,
   ybar,
   bar width=10pt,
   xlabel={Year},
   ylabel={Number of Devices (Billions)},
   ymin=0,
   ymax=95, % a little headroom so labels don't crowd
   xtick=data,
   xticklabel style={rotate=45, anchor=east, font=\normalsize},
   yticklabel style={font=\normalsize},
   nodes near coords,
   nodes near coords align={vertical},
   every node near coord/.append style={
        font=\fontsize{6}{7}\selectfont, % slightly smaller than \tiny
        yshift=3pt % lift labels so they don't overlap
   },
]
\addplot[
    fill=blue!60,
] coordinates {
    (2015,15.41)
    (2016,17.7)
    (2017,20.4)
    (2018,23.1)
    (2019,26.7)
    (2020,30.7)
    (2021,35.8)
    (2022,42.6)
    (2023,51.1)
    (2024,62.1)
    (2025,75.44)
};
\end{axis}
\end{tikzpicture}
\caption{Growth of global IoT-connected devices (2015--2025). Data derived from~\cite{9522283}.}
\label{fig:iot_growth}
\end{figure}

According to \cite{9522283}, the number of Internet-connected IoT devices increased from approximately 15.41 billion in 2015 to nearly 75.44 billion by 2025, demonstrating the rapid and extensive adoption of IoT technologies worldwide. Fig.~\ref{fig:iot_growth}, shows this continued upward growth, highlighting the scale and acceleration of IoT expansion worldwide.

\section{Limitations of Existing ML-Based IDS in IoT Environments}
IoT environments introduce unique security challenges as compromised devices can serve as  potential entry points into broader networks, enabling attackers to perform lateral movement across networked systems and privilege escalation (gaining admin/root access). In addition, communication channels between IoT devices, gateways, and backend services are susceptible to interception, spoofing, and manipulation, particularly in resource-constrained and unmanaged deployment scenarios\cite{app11104580}.

Despite the widespread adoption of machine learning (ML) techniques for IoT intrusion detection, existing research lacks a unified framework that combines detection effectiveness and deployment feasibility under real-world operational constraints. Most prior studies emphasize high detection accuracy while ignoring critical factors such as computational complexity, memory consumption, inference latency, and architectural suitability across edge, fog/gateway, and cloud layers. Consequently, many proposed ML-based intrusion detection solutions remain difficult to deploy in practical IoT environments, limiting their real-time effectiveness and deployment scalability.

\section{DEPLOYMENT CRITERIA AND ARCHITECTURE LAYERS}
To enable a structured and reproducible deployment-aware evaluation, this study defines representative quantitative deployment capability envelopes for edge, fog/gateway, and cloud layers. Unlike prior work that describes architectural layers qualitatively, the capability envelopes used in this study are grounded in hardware specifications, experimental benchmarking platforms, and system-level modeling parameters reported in the literature. These values are used to characterize relative constraint dominance across layers rather than to imply universal hardware guarantees.

\subsection{Edge Layer (IoT Devices)}\label{AA}
The edge layer consists of highly resource-constrained IoT devices such as embedded sensors, controllers, and smart endpoints. Representative experimental platforms report CPU frequencies ranging from 1.43–2.26 GHz and AI acceleration capacities between 4–40 tera operations per second (TOPS) \cite{Pham_2024}. RAM configurations typically range from 4–32 GB, with measured runtime usage between 2.61–3.11 GB under anomaly detection workloads \cite{Pham_2024}. Power budgets are limited, with standard CPU consumption reported between 0.9–3.4 W and operating modes extending up to 5–15 W depending on configuration \cite{Pham_2024}, \cite{GarciaPerez2023EdgeAI}. Latency-sensitive applications often impose network delay requirements below 10 ms, with augmented-reality scenarios requiring response times below 50 ms \cite{8567673}.

These representative platforms include both lightweight IoT endpoints and modern AI-enabled edge computing devices that integrate dedicated accelerators. While such edge platforms may provide relatively high computational capability, they remain significantly more constrained than centralized cloud infrastructures and must operate under strict energy and latency limitations.

Overall, these characteristics constraints indicate strong computational and memory scarcity, strict energy limitations, and ultra-low-latency requirements, making lightweight models preferable for edge deployment.

\subsection{Fog / Gateway Layer}

The fog or gateway layer aggregates and processes traffic from multiple edge devices. Representative studies report multi-core CPUs operating between 1.2–1.8 GHz and RAM capacities ranging from 1–16 GB\cite{8241101}. Network characteristics vary depending on communication technologies, including 45 kbps Long Range (LoRa), 8 Mbps (WiFi), and 25 Mbps (4G) links \cite{8241101}. Compared to edge devices, fog nodes provide moderate computational and memory resources while maintaining low-latency processing suitable for time-sensitive aggregation and intermediate analytics.

These characteristics position fog environments as a balanced layer capable of supporting moderately complex ML models that exceed the capacity of individual IoT devices.

\subsection{Cloud Layer}
The cloud layer provides centralized and elastic computational infrastructure. System-level modeling studies report edge-cloud architectures consisting of racks with five processors operating at 3.33 Giga Instructions Per Second (GIPS) each and large-scale deployments reaching approximately 5.5K processors \cite{8567673}. Network bandwidth between edge computing nodes can reach 1–100 Gbps, while response times typically range from about 30 ms under light load to 170 ms under heavier system load. \cite{8567673}. Such infrastructure supports high-memory environments, GPU acceleration, and large-scale parallel computation.

Compared to edge and fog layers, cloud platforms exhibit minimal computational resource constraints but may introduce additional communication latency and greater dependence on network connectivity. Consequently, computationally intensive deep learning and optimization-driven models are most suitable for cloud deployment.

The representative quantitative capability envelopes summarized in Table I characterize the relative constraint dominance across edge, fog/gateway, and cloud layers. These values are extracted from experimental platforms and system-level modeling studies and are intended to represent typical capability ranges rather than universal hardware specifications. They provide the foundation for the structured deployment feasibility framework introduced in the following section.

\begin{table*}[htbp]
\caption{Representative Quantitative Deployment Constraints Extracted from Prior Studies}
\centering
\resizebox{\textwidth}{!}{
\begin{tabular}{|p{2.5cm}|p{3.5cm}|p{4cm}|p{4cm}|p{2cm}|}
\hline
\textbf{Layer} & 
\textbf{Representative Compute Capability} & 
\textbf{Representative Memory / Power} & 
\textbf{Representative Network / Latency} & 
\textbf{Source(s)} \\ 
\hline

\textbf{Edge (IoT Devices)} &
1.43--2.26 GHz CPUs; 4--40 TOPs &
4--32 GB RAM; runtime usage 2.61--3.11 GB; 0.9--3.4 W (standard); 5--15 W power modes &
Network delay $<$ 10 ms; AR latency requirement $<$ 50 ms &
{\cite{Pham_2024}, \cite{8567673}, \cite{GarciaPerez2023EdgeAI}} \\ 
\hline

\textbf{Fog / Gateway} &
1.2--1.8 GHz multi-core CPUs; CPU capacity 100\%--800\% &
1--16 GB RAM &
45 kbps (LoRa); 8 Mbps (WiFi); 25 Mbps (4G) &
{\cite{8241101}} \\ 
\hline

\textbf{Cloud} &
5 processors @ 3.33 GIPS per rack; 5.5K processors total &
Multi-rack infrastructure &
1--100 Gbps inter-edge bandwidth; 30 ms (Load=1) to 170 ms (Load=10) response time &
{\cite{8567673}} \\ 
\hline

\end{tabular}}
\label{tab:deployment_constraints}
\end{table*}

\section{DEPLOYMENT FEASIBILITY SCORING MODEL}

Building upon the quantitative capability envelopes defined in
Section III, this section introduces the Deployment Feasibility
Score (DFS) framework for systematically evaluating the suitability
of machine learning–based IDS models across edge, fog/gateway,
and cloud deployment layers.

To align model resource requirements with deployment-layer constraints, the proposed DFS framework provides a lightweight ordinal scoring mechanism that formalizes deployment suitability across heterogeneous computing environments.

\subsection{Resource Normalization and Ordinal Scoring }
To ensure reproducibility in the assignment of ordinal scores, the proposed framework converts model resource requirements into normalized demand indicators relative to the capability limits of the target deployment platform.

Three deployment-relevant metrics are considered: computational demand, memory footprint, and inference latency sensitivity. Each metric is mapped onto a five-level ordinal scale representing increasing resource demand.\\

The normalized scores are computed as follows:

\begin{equation}
C = \min\left(5,\left\lceil 5 \times \frac{\mathrm{Ops}_{\mathrm{model}}}{\mathrm{Ops}_{\max}} \right\rceil\right)
\end{equation}

\begin{equation}
M = \min\left(5,\left\lceil 5 \times \frac{\mathrm{Mem}_{\mathrm{model}}}{\mathrm{Mem}_{\max}} \right\rceil\right)
\end{equation}

\begin{equation}
L = \min\left(5,\left\lceil 5 \times \frac{\mathrm{Latency}_{\mathrm{model}}}{\mathrm{Latency}_{\max}} \right\rceil\right)
\end{equation}

where $\mathrm{Ops}_{\mathrm{model}}$ represents the number of operations required for model inference, $\mathrm{Mem}_{\mathrm{model}}$ represents runtime memory consumption, and $\mathrm{Latency}_{\mathrm{model}}$ represents measured inference latency. Similarly, $\mathrm{Ops}_{\max}$, $\mathrm{Mem}_{\max}$, and $\mathrm{Latency}_{\max}$ denote representative capability limits of the target deployment platform.

Ratios $Ops_{model}/Ops_{max}$, $Mem_{model}/Mem_{max}$, and
$Latency_{model}/Latency_{max}$ represent the fractions of the available computational capacity, memory resources, and latency budget of the target device consumed by a single model inference.

Values approaching unity indicate 
that the model nearly saturates the corresponding system resource. In practical 
edge computing environments, devices must also support operating system 
services, network communication, sensor data acquisition, and other concurrent 
applications. Consequently, models that operate close to the resource limits of 
the device may lead to resource contention, increased response latency, and 
reduced system stability. In such cases, deployment may require model 
optimization or migration to a higher-capability platform such as a fog gateway 
or cloud infrastructure.

The ceiling operator $\lceil \cdot \rceil$ ensures that fractional normalized values are rounded upward to the nearest integer, thereby conservatively reflecting the computational demand of the model and preventing underestimation of resource requirements.

\subsection{Deployment Feasibility Score Formulation    }

The scaling factor of 5 maps normalized resource utilization onto the five-level ordinal scoring scale defined in Table~II. The ceiling operator $\lceil \cdot \rceil$ ensures conservative classification by assigning models to the next higher demand level once resource usage exceeds the lower bound of a category. The outer minimum operator guarantees that the resulting scores remain bounded within the defined ordinal range.

This normalization strategy ensures that deployment feasibility evaluation remains consistent, reproducible, and platform-aware, thereby enabling fair comparison of heterogeneous machine learning models across edge, fog, and cloud environments.

The deployment feasibility score (DFS) is defined as:

\begin{equation}
\text{DFS} = w_c C + w_m M + w_l L
\end{equation}

where:

\begin{itemize}
\item $C$ denotes the computational demand of the model.
\item $M$ denotes the memory footprint required for deployment.
\item $L$ denotes the inference latency sensitivity.
\item $w_c$, $w_m$, and $w_l$ denote layer-specific weights that capture the relative importance of compute, memory, and latency constraints across different deployment layers.
\end{itemize}

The weights satisfy:
\begin{equation}
w_c + w_m + w_l = 1
\end{equation}

This ensures normalized contribution across evaluation dimensions.

The linear weighted formulation is selected for its interpretability
and transparency in deployment-oriented decision analysis.
Unlike complex multi-objective optimization approaches, the weighted
aggregation enables straightforward comparison across heterogeneous
machine learning models while maintaining computational simplicity
suitable for rapid deployment feasibility evaluation.

Each component is assigned on a 1--5 ordinal scale,
reflecting increasing resource demand, as summarized
in Table~II.

\begin{table}[t]
\caption{Ordinal Scoring Interpretation}
\centering
\begin{tabular}{|c|l|}
\hline
\textbf{Score} & \textbf{Interpretation} \\ 
\hline
1 & Very Low Demand \\ 
2 & Low Demand \\ 
3 & Moderate Demand \\ 
4 & High Demand \\ 
5 & Very High Demand \\ 
\hline
\end{tabular}
\label{tab:scoring_scale}
\end{table}

\begin{enumerate}
\item \textbf{Computational Demand ($C$)}: 
Computational demand reflects model structural complexity and inference cost. Linear and probabilistic classifiers (e.g., Logistic Regression, Naïve Bayes) are assigned lower scores (1--2), tree-based ensembles receive moderate scores (3), while deep neural networks, recurrent architectures, and optimization-driven pipelines receive higher scores (4--5) due to increased processing requirements\cite{diab2025intrusiondetectionresourceconstrainediot}.

\item \textbf{Memory Demand ($M$)}: 
Memory demand reflects the memory resources required by the model, including parameter storage, model size, and runtime memory footprint. Lightweight models such as single-tree classifiers and linear models are assigned lower scores (1–2). Moderate memory usage models, including tree-based ensembles and clustering-based systems, receive intermediate scores (3), while deep neural networks and large ensemble architectures receive higher scores (4–5) due to increased parameter storage and activation memory requirements.

\item \textbf{Latency Sensitivity ($L$)}: 
Latency sensitivity represents tolerance to inference delay in operational deployment. Models intended for real-time detection at the device level receive lower scores (1–2), reflecting strict latency requirements. Applications with moderate latency tolerance, such as gateway-level monitoring, receive intermediate scores (3), while batch-processing or centralized analytics systems receive higher scores (4–5), where longer processing delays are acceptable.
\end{enumerate}

\subsection{Layer-Specific Weight Selection}

Weights are derived from the relative constraint dominance observed in the quantitative capability envelopes summarized in Table~\ref{tab:deployment_constraints}. Because edge devices exhibit strict computational and energy limitations, fog nodes balance processing and latency requirements, and cloud platforms prioritize large-scale computation with relaxed real-time constraints, weight assignments differ across layers.

\textit{Edge Layer (Resource-Constrained, Power-Limited):} 
\begin{equation}
w_c = 0.40, \quad w_m = 0.35, \quad w_l = 0.25
\end{equation}
Computational and memory scarcity dominate feasibility at the edge, while latency—although strict—is comparatively uniform across deployed models.

\textit{Fog / Gateway Layer (Near-Real-Time Aggregation):} 
\begin{equation}
w_c = 0.35, \quad w_m = 0.30, \quad w_l = 0.35
\end{equation}
Fog environments balance computational availability with latency sensitivity due to aggregation and intermediate analytics roles.

\textit{Cloud Layer (Elastic Centralized Infrastructure):} 
\begin{equation}
w_c = 0.25, \quad w_m = 0.25, \quad w_l = 0.50
\end{equation}
Cloud platforms offer elastic computational and memory resources; therefore, latency and architectural placement dominate deployment considerations.

These weights are policy-driven and reflect the relative importance of computational, memory, and latency constraints at each architectural layer. They are selected based on typical resource limitations of edge, fog, and cloud environments rather than being derived through empirical optimization.

\subsection{Feasibility Classification}

The resulting DFS is mapped to qualitative feasibility categories as shown in Table~\ref{tab:dfs_classification}.

\begin{table}[htbp]
\caption{DFS-Based Feasibility Classification}
\centering
\begin{tabular}{|c|l|}
\hline
\textbf{DFS Range} & \textbf{Feasibility Level} \\ 
\hline
1.0--2.0 & High Feasibility \\ 
2.1--3.5 & Moderate Feasibility \\ 
3.6--5.0 & Low Feasibility \\ 
\hline
\end{tabular}
\label{tab:dfs_classification}
\end{table}

Lower DFS values indicate stronger deployment suitability within the evaluated layer.

\subsection{Example Application}

To illustrate the framework, three representative models are evaluated: Naïve Bayes, Random Forest, and Convolutional Neural Networks (CNN). The assigned ordinal scores represent typical demand characteristics reported in the literature and are used here for illustrative evaluation of the DFS framework.

The following example demonstrates how the proposed Deployment Feasibility Score (DFS) framework can be applied to representative machine learning model families commonly used in intrusion detection systems.

\begin{table}[htbp]
\caption{Illustrative DFS Evaluation Across Layers}
\centering
\footnotesize
\setlength{\tabcolsep}{4pt}
\begin{tabular}{|l|c|c|c|c|c|c|}
\hline
\textbf{Model} & \textbf{C} & \textbf{M} & \textbf{L} & \textbf{Edge DFS} & \textbf{Fog DFS} & \textbf{Cloud DFS} \\
\hline
Na\"{\i}ve Bayes & 1 & 1 & 1 & 1.0 & 1.0 & 1.0 \\
Random Forest & 3 & 3 & 3 & 3.0 & 3.0 & 3.0 \\
CNN & 5 & 4 & 4 & 4.40 & 4.35 & 4.25 \\
\hline
\end{tabular}
\label{tab:dfs_example}
\end{table}

For example, Random Forest under edge-layer weights is computed as:

\begin{equation}
\text{DFS} = (0.40)(3) + (0.35)(3) + (0.25)(3) = 3.0
\end{equation}

Resulting in a DFS of 3.0, which corresponds to moderate feasibility at the edge layer as defined in Table III.

In contrast, CNN models show lower feasibility for edge deployment due to higher computational and memory requirements. Although feasibility improves slightly in cloud environments with greater computational resources, the overall DFS remains relatively high compared with lightweight statistical or tree-based models.

\subsection{Sensitivity Analysis}
To evaluate the robustness of the proposed Deployment Feasibility Score (DFS) framework, a sensitivity analysis was conducted by varying the layer-specific weight assignments while keeping the normalization constraint $w_c + w_m + w_l = 1$.

For the edge layer, the baseline weight configuration $(0.40, 0.35, 0.25)$ was adjusted to two alternative configurations:

\begin{itemize}
\item Scenario A: $(0.35, 0.35, 0.30)$
\item Scenario B: $(0.45, 0.30, 0.25)$
\end{itemize}

These variations reflect moderate shifts in the dominance of computational, memory, and latency constraints while remaining architecturally realistic.

The Deployment Feasibility Score (DFS) was recomputed for representative models—Naïve Bayes, Random Forest, and Convolutional Neural Networks (CNN)—with the results summarized in Table ~\ref{tab:sensitivity}.

\begin{table}[h]
\caption{Sensitivity Analysis of DFS Under Varying Edge Weights}
\label{tab:sensitivity}
\centering
\begin{tabular}{lccc}
\hline
Model & Baseline DFS & Scenario A & Scenario B \\
\hline
Naïve Bayes & 1.00 & 1.00 & 1.00 \\
Random Forest & 3.00 & 3.00 & 3.00 \\
CNN & 4.45 & 4.35 & 4.50 \\
\hline
\end{tabular}
\end{table}

As shown in Table~\ref{tab:sensitivity}, while absolute DFS values exhibit minor numerical variation under alternative weight configurations, the relative ranking of models remains unchanged. Naïve Bayes consistently demonstrates high feasibility for edge deployment, Random Forest maintains moderate feasibility, and CNN remains less suitable for resource-constrained environments. The stability of model ordering under moderate weight modifications confirms that the DFS framework is not overly sensitive to policy-driven parameter selection. This robustness strengthens the validity of the deployment-aware evaluation mechanism.

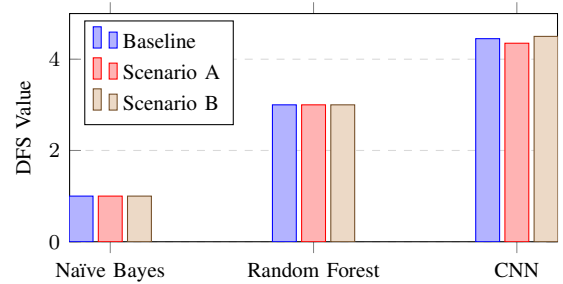
\begin{figure}[h]
\centering
\begin{tikzpicture}
\begin{axis}[
    width=\columnwidth,
    height=4.6cm,
    ybar,
    bar width=9pt,
    ylabel={DFS Value},
    symbolic x coords={Na\"{\i}ve Bayes, Random Forest, CNN},
    xtick=data,
    ymin=0, ymax=5,
    ymajorgrids=true,
    grid style={dashed,gray!30},
    legend pos=north west,
    legend cell align=left,
    tick label style={font=\footnotesize},
    label style={font=\footnotesize},
    legend style={font=\footnotesize},
]
\addplot coordinates {(Na\"{\i}ve Bayes,1.0) (Random Forest,3.0) (CNN,4.45)};
\addlegendentry{Baseline}
\addplot coordinates {(Na\"{\i}ve Bayes,1.0) (Random Forest,3.0) (CNN,4.35)};
\addlegendentry{Scenario A}
\addplot coordinates {(Na\"{\i}ve Bayes,1.0) (Random Forest,3.0) (CNN,4.50)};
\addlegendentry{Scenario B}
\end{axis}
\end{tikzpicture}
\caption{DFS Stability Under Weight Perturbation}
\label{fig:sensitivity}
\end{figure}

\subsection{Formal Justification of the Modeling Design}

The proposed Deployment Feasibility Score (DFS) framework uses a simple 1–5 scale to represent computational demand, memory usage, and latency requirements. Instead of relying on exact hardware measurements, this scale compares models based on their relative resource needs. Because IoT, fog, and cloud systems vary widely in hardware capabilities and continue to evolve rapidly, fixed numerical benchmarks may not apply to all platforms. Therefore, using an ordinal scale allows for a consistent, clear, and reproducible way to compare model suitability across different architectural layers. 

The linear weighted aggregation model is used to combine these factors in a straightforward and interpretable manner. The model is defined as:
\begin{equation}
DFS = w_c C + w_m M + w_l L
\end{equation}

This study does not aim to perform detailed multi-criteria optimization. Instead, it provides a structured screening mechanism that connects model resource demands to the constraints of different architectural layers. A linear formulation is adopted for its interpretability, ease of sensitivity analysis, and support for policy-driven weight adaptation.

Instead of employing complex ranking algorithms, this study adopts a simple scoring formulation that provides reliable results while remaining easier for researchers and practitioners to apply.

\subsection{Comparison With Existing Surveys and Positioning of This Work}

Prior surveys have extensively reviewed machine learning (ML) and deep learning (DL) approaches for intrusion detection in IoT environments, focusing on taxonomies of IDS techniques, benchmark datasets, attack classifications, and evaluation metrics. For instance, Rahman \emph{et al.} provide a comprehensive overview of IDS strategies in IoT networks, emphasizing datasets, detection metrics, and practical considerations such as computational efficiency and lightweight operation on constrained devices \cite{RAHMAN2025100082}. Mohammed and Talib similarly review ML algorithms for IDS and highlight core IoT challenges including limited computational capacity, real-time constraints, and the need for lightweight models \cite{Mohammed2024IoTIDS}.

Other surveys concentrate on specific methodological families. Gueriani \emph{et al.} analyze deep reinforcement learning (DRL)–based intrusion detection in IoT settings, covering taxonomies, datasets, and performance metrics \cite{Gueriani_2023}. Broader IoT security surveys summarize the adoption of ML/DL techniques while identifying recurring limitations such as resource constraints, scalability issues, and deployment practicality \cite{Harahsheh2023IoTSurvey}. Earlier IoT-focused IDS reviews present ML-driven classifications and discuss anomaly-based, signature-based, and specification-based detection approaches, yet primarily evaluate techniques through detection-centric perspectives \cite{Mliki2021IoTIDS}. More recent comprehensive IDS surveys further emphasize operational challenges such as high false positive rates and elevated computational requirements that restrict deployment in resource-constrained IoT environments \cite{Hozouri2025IDS}.

Although prior surveys provide valuable insights, they do not offer a unified method for quantitatively evaluating deployment feasibility across different architectural layers. They rarely connect model resource demands to the actual constraints of edge, fog, and cloud environments. Therefore, some IDS models that achieve high accuracy in research settings may not be suitable for real-world IoT deployment.

To address this gap, this paper introduces a lightweight Deployment Feasibility Score (DFS) framework that complements detection effectiveness by enabling structured, layer-aware feasibility comparison across edge, fog/gateway, and cloud environments. Table~\ref{tab:survey_comp} summarizes the key distinctions between prior surveys and the proposed deployment-aware framework.

\begin{table*}[t]
\centering
\caption{Comparison of Representative IoT IDS Surveys and the Proposed Deployment-Aware Framework}
\label{tab:survey_comp}
\renewcommand{\arraystretch}{1.15}
\begin{tabular}{p{3.2cm} p{1cm} p{1.6cm} p{1.6cm} p{1.8cm} p{2.2cm}}
\hline
\textbf{Reference} & \textbf{Year} & \textbf{IDS/ML Taxonomy} & \textbf{Datasets \& Metrics} & \textbf{Resource Constraint Discussion} & \textbf{Quantitative Deployment Modeling (Edge/Fog/Cloud)} \\
\hline
Rahman \emph{et al.} \cite{RAHMAN2025100082} & 2025 & \checkmark & \checkmark & \checkmark & \texttimes \\
Mohammed \& Talib \cite{Mohammed2024IoTIDS} & 2024 & \checkmark & \checkmark & \checkmark & \texttimes \\
Gueriani \emph{et al.} \cite{Gueriani_2023} & 2023 & \checkmark & \checkmark & \checkmark (general) & \texttimes \\
Harahsheh \& Chen 
\cite{Harahsheh2023IoTSurvey} & 2023 & \checkmark & \checkmark & \checkmark & \texttimes \\
Mliki \emph{et al.} \cite{Mliki2021IoTIDS} & 2021 & \checkmark & \checkmark & \checkmark & \texttimes \\
Hozouri \emph{et al.} \cite{Hozouri2025IDS} & 2025 & \checkmark & \checkmark & \checkmark & \texttimes \\
\hline
\textbf{This Work} & -- & \checkmark & \checkmark & \checkmark & \checkmark \\
\hline
\end{tabular}
\end{table*}

\section{Deployment-Aware Analysis of ML-Based IDS Techniques}
\subsection{DFS-Based Evaluation of Representative IDS Models}
Building upon the comparative analysis presented in the previous section, this section applies the proposed Deployment Feasibility Score (DFS) framework to representative machine learning–based IDS techniques. Unlike traditional evaluations that rely primarily on detection accuracy, the DFS-based approach systematically incorporates computational complexity, memory requirements, and latency sensitivity within layer-specific architectural constraints.

Using the quantitative capability  defined in Table I and the ordinal scoring method introduced in Section IV, representative models are evaluated across edge, fog/gateway, and cloud environments. This structured deployment-aware analysis shows that model suitability can vary significantly depending on the architectural context, even when detection performance is similar.
\subsection{Empirical Benchmark Validation Using Published Deployments}

To complement the conceptual Deployment Feasibility Score (DFS) analysis, we examine empirical deployment evidence reported in prior work. The goal is to validate whether the feasibility trends predicted by the DFS framework align with measured performance of machine learning and deep learning intrusion detection models deployed on resource-constrained IoT edge platforms.

A recent hardware-aware intrusion detection study implemented several ML and DL models on a Raspberry Pi 3 Model B+, a widely used IoT gateway platform equipped with a quad-core 1.4 GHz ARM Cortex-A53 processor and 1 GB RAM. The study evaluates model accuracy, memory footprint, computational complexity, and on-device inference efficiency for intrusion detection tasks using the Edge-IIoTset benchmark dataset
\cite{diab2025intrusiondetectionresourceconstrainediot}.

The reported results demonstrate clear differences in computational demand and deployment efficiency across model families. Lightweight tree-based models such as LightGBM and XGBoost achieve strong detection performance while maintaining extremely small memory footprints and low inference latency. For example, LightGBM reaches 95.25\% accuracy while requiring only 74.93 KB of flash and 1.13 KB of RAM, with an average inference latency of 27 ms and energy consumption of 6.75 mJ per prediction on the Raspberry Pi device
\cite{diab2025intrusiondetectionresourceconstrainediot}.

Similarly, XGBoost achieves 95.11\% accuracy with slightly higher memory requirements but even faster inference latency of 22 ms. Random Forest models achieve comparable accuracy (94.12\%) but exhibit higher RAM usage and slower inference time due to deeper tree traversal operations
\cite{diab2025intrusiondetectionresourceconstrainediot}.

In contrast, deep learning models provide improved detection accuracy but incur significantly higher computational costs. The optimized 1D-CNN intrusion detection model achieves 96.73\% accuracy, but requires substantially higher computational complexity (838.89 K FLOPs) and longer inference latency (300 ms) compared with tree-based models\cite{diab2025intrusiondetectionresourceconstrainediot}.

These empirical results highlight an important trade-off between detection accuracy and deployment efficiency. Tree-based models provide efficient real-time detection with minimal memory and energy consumption, making them highly suitable for deployment directly on edge devices. Deep learning architectures offer improved detection capability but introduce higher computational overhead and latency, suggesting that they may be better suited for environments where additional computational resources are available.

Table VII summarizes the reported deployment results and illustrates how the observed hardware characteristics align with the feasibility trends predicted by the DFS framework. Lightweight models with low computational demand and memory footprint exhibit stronger feasibility for edge deployment, whereas deep learning models with higher computational complexity incur greater deployment cost.

\begin{table*}[t]
\centering
\caption{Empirical Deployment Performance of IDS Models on Raspberry Pi Edge Gateway}
\label{tab:empirical_deployment}
\renewcommand{\arraystretch}{1.2}

\resizebox{\textwidth}{!}{
\begin{tabular}{|l|l|c|c|p{3.0cm}|p{2.8cm}|p{3.0cm}|}
\hline
\textbf{Model} & \textbf{Platform} & \textbf{Accuracy} & \textbf{F1 Score} & \textbf{Model Footprint} & \textbf{On-Device Inference} & \textbf{Deployment Implication} \\
\hline
LightGBM & Raspberry Pi 3 B+ & 95.25\% & 94.74\% & 74.93 KB flash, 1.13 KB RAM, 1.20 K Ops & 27 ms, 250 mW, 6.75 mJ & Highly suitable for edge deployment \\
\hline
XGBoost & Raspberry Pi 3 B+ & 95.11\% & 94.42\% & 266.59 KB flash, 0.51 KB RAM, 4.27 K Ops & 22 ms, 250 mW, 5.50 mJ & Efficient edge inference \\
\hline
Random Forest & Raspberry Pi 3 B+ & 94.12\% & 93.21\% & 211.22 KB flash, 4.61 KB RAM, 3.38 K Ops & 70 ms, 325 mW, 22.75 mJ & Edge / gateway deployment feasible \\
\hline
1D-CNN & Raspberry Pi 3 B+ & 96.73\% & 97.24\% & 190.34 KB flash, 6.89 KB RAM, 838.89 K FLOPs & 300 ms, 275 mW, 82.50 mJ & Higher computational cost; better suited for gateway or fog deployment \\
\hline
\end{tabular}
}

\end{table*}

\section{Limitations and Discussion}

While the proposed Deployment Feasibility Score (DFS) framework provides a structured mechanism for evaluating machine learning (ML)–based intrusion detection techniques across edge, fog/gateway, and cloud architectures, several limitations must be acknowledged.

First, the DFS model uses a 1–5 ordinal scale to represent computational demand, memory usage, and latency sensitivity. While this approach supports consistent comparison and avoids false numerical precision, it does not capture detailed runtime differences that may arise from specific hardware platforms or optimization techniques.

Second, the weights assigned to each layer are policy-driven and based on representative hardware capabilities from prior studies. While these values capture general constraint patterns, they do not represent every possible deployment scenario. Factors such as processor type, hardware acceleration, model compression, or distributed inference can influence real-world feasibility.

Third, the evaluation in this study focuses on representative ML models rather than conducting large-scale empirical benchmarking. The goal of this work is to formalize a deployment-aware feasibility framework, not to optimize detection performance metrics such as accuracy or F1-score. Future research may incorporate empirical runtime measurements to further validate the DFS framework under controlled experimental conditions.

Despite these limitations, the proposed DFS framework advances deployment-aware evaluation by systematically linking model resource demands to architectural constraints. By moving beyond accuracy-centric analysis, this work provides a reproducible and adaptable feasibility modeling approach that supports informed decision-making across heterogeneous IoT infrastructures. The framework thus contributes a practical foundation for aligning machine learning–based intrusion detection with real-world deployment realities.

Table VII summarizes the empirical deployment performance of representative IDS models on a Raspberry Pi 3 Model B+ edge gateway. Both accuracy and F1-score indicate strong detection capability for lightweight tree-based models, which also maintain low latency and energy consumption suitable for edge deployment. The 1D-CNN achieves the highest F1-score but introduces higher computational cost, suggesting better suitability for gateway or fog environments.

\section{Conclusion}

This paper presented a deployment-aware evaluation framework for machine learning–based intrusion detection systems in Internet of Things (IoT) environments. Unlike previous surveys that focus mainly on detection accuracy or algorithm categories, the proposed Deployment Feasibility Score (DFS) evaluates whether a model is practical for real-world deployment. It considers computational demand, memory usage, and latency requirements across edge, fog/gateway, and cloud architectures.

By using a simple, normalized scoring method, the DFS framework enables clear and structured comparison of ML models under realistic architectural constraints. The results show that lightweight statistical and linear models are best suited for resource-limited edge devices, ensemble methods are more appropriate for fog/gateway environments, and deep learning models are most feasible in cloud infrastructures.

Overall, the proposed DFS framework introduces a reproducible and architecture-aware evaluation mechanism that bridges the gap between detection performance and deployment practicality, contributing a structured methodology for real-world IoT security system design.

\bibliographystyle{IEEEtran}
\bibliography{references}
\end{document}